\documentclass[twocolumn,showpacs,amsmath,floatfix]{revtex4}
\usepackage{graphicx}

\pdfoutput=1

\begin{document}

\title{A new test of local Lorentz invariance using $^{21}$Ne-Rb-K comagnetometer}
\author{M. Smiciklas}
\author{J. M. Brown}
\author{L.W. Cheuk}
\author{M. V. Romalis}

\begin{abstract}
We develop a new comagnetometer using $^{21}$Ne atoms with nuclear spin $I=3/2$
and Rb atoms polarized by spin-exchange with K atoms to search for tensor
interactions that violate local Lorentz invariance. We frequently reverse
orientation of the experiment and search for signals at the first and second harmonics of
the sidereal frequency. We constrain 4 of the 5 spatial
Lorentz-violating coefficients $c^n_{jk}$ that parameterize anisotropy of the
maximum attainable velocity of a neutron at a level of $10^{-29}$, improving
previous limits by 2 to 4 orders of magnitude and placing the most stringent
constrain on deviations from local Lorentz invariance.
\end{abstract}

\affiliation{Department of Physics, Princeton University, Princeton, New Jersey 08544}
\pacs{11.30.Cp, 21.30.Cb, 32.30.Dx}
\maketitle

The Michelson-Morley experiment and its successors have established that the speed of light is isotropic
to a part in $10^{17}$ \cite{Schiller,Peters}. Similarly, possible anisotropy in the maximum attainable
velocity (MAV) for a massive particle \cite{Coleman} has been constrained by  Hughes and Drever
NMR experiments \cite{Hughes, Drever} and their successors to a part in $10^{27}$ \cite{clockcom}. These experiments
form the basis for the principle of local Lorentz invariance (LLI). Together with the weak equivalence principle
and the position invariance principle, they constitute the Einstein equivalence principle that is the basis of general relativity \cite{Will}.
Measurements of tensor NMR energy shifts \cite{Lamoreaux,Chupp} are
particularly sensitive to variation in MAV due to a finite kinetic energy of valence nucleons. They place the most stringent limits
on violation of LLI within the $TH\epsilon \mu $ formalism \cite{Lightman} describing
deviations from the Einstein Equivalence Principle as well as within more general Standard Model Extension (SME) \cite{Kost}.
They compare favorably even to  the limits on variation in MAV from ultra-high energy cosmic rays and other astrophysical phenomena \cite{Altschul,Altschul1,Scully}.
It can be argued that Lorentz invariance is likely to be broken at some level by the effects
of quantum gravity, which contains a dimensionfull Planck scale that is not
Lorentz-invariant. Popular ideas for quantum gravity theories, such as recently proposed
Ho\v{r}ava-Lifshitz model \cite{Horava}, explicitly violate Lorentz symmetry.  CPT-even tensor Lorentz-violating effects, such as variation in MAV, are
particularly interesting to explore because they can arise from purely kinematic violation of Lorentz invariance, do not require explicit particle spin coupling at the fundamental level, and do not suffer from fine-tuning problems associated with CPT-odd Lorentz-violating vector spin interactions \cite{Mattingly,Posp3}.

Here we describe a new comagnetometer that is sensitive to anisotropy in neutron MAV at $10^{-29}$ level.
The idea of the experiment is based on the K-$^{3}$He comagnetometer, previously used to constrain Lorentz-violating vector spin
interactions \cite{Brown}. The $^{3}$He ($I=1/2$) is replaced by $^{21}$Ne
 ($I=3/2$) to allow measurements of tensor anisotropy. In addition, since the gyromagnetic ratio of $^{21}$Ne is about
an order of magnitude smaller than that of $^{3}$He, the comagnetometer has
an order of magnitude better energy resolution for the same level of
magnetic field sensitivity. The electric quadrupole interactions of $^{21}$Ne cause several problems. In order to overcome a faster nuclear spin
relaxation in $^{21}$Ne relative to $^{3}$He, we replace K
with Rb atoms, which have a larger spin-exchange cross-section with $^{21}$Ne \cite{Ghosh}.  We also increase the alkali density by on order
of magnitude and rely on hybrid optical pumping \cite{hybrid}, using
spin-exchange with optically-pumped K to polarize the Rb atoms while avoiding strong absorption of pumping light by the optically dense
Rb vapor.

The experiment is placed on a rotary platform and its orientation can be
frequently reversed to introduce a modulation of the Lorentz-violating signal (Fig.~1).
Among lab-fixed backgrounds, only the gyroscopic signal due to the Earth's rotation
cannot be easily suppressed, and we rely on sidereal and semi-sidereal
oscillations expected for Lorentz-violating effects to extract anisotropic
signals of an extra-solar origin. We measure the $^{21}$Ne NMR frequency
with a sensitivity of about $0.5~$nHz and, within the Schmidt model for the $^{21}$Ne nucleus, constrain  4 of the 5
spatial components of the symmetric traceless SME tensor $c^n_{jk}$ for neutrons at a level of
a few parts in $10^{29}$. This sensitivity exceeds all other limits on the $c_{\mu \nu}$ coefficients in the matter sector \cite{Kosttable} as well as the recent
limit on the neutrino $c_{0j}$ coefficient at a level of $10^{-27}$ \cite{Icecube} and limits on $c_{00}$ at a level of $10^{-23}$  from analysis
of ultra-high energy cosmic rays \cite{Scully,Bi}.

\begin{figure}[b]
\centering
\includegraphics[width=6cm]{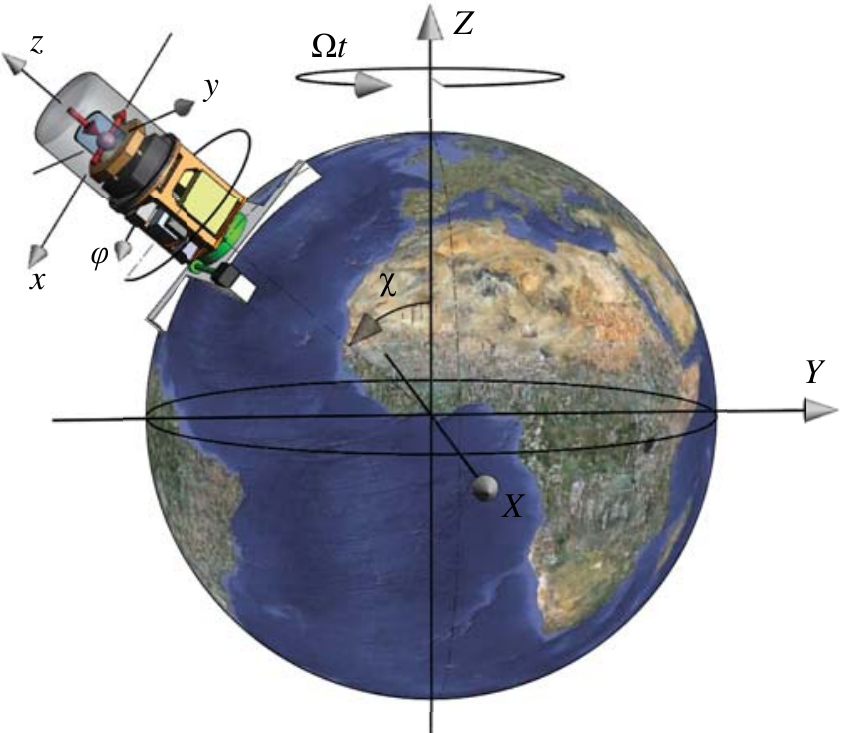}
\caption{The experimental apparatus is rotated around the local vertical. $^{21}$Ne spins are polarized down along $-\hat{z}$ and the probe beam is directed horizontally along $-\hat{x}$}.\label{fig_setup}
\end{figure}

The operating principles of  the comagnetometer are similar to that
described in \cite{gyro,Tomthesis,Brown}.  The atoms are contained in a
1.4 cm diameter spherical cell made from aluminosilicate glass that is
filled with Ne (enriched to 95\% of $^{21}$Ne) at a density of $2.03\pm 0.05$ amagat as determined from
pressure broadening of Rb D1 line \cite{Perram}, 30 Torr of N$_{2}$ for quenching, and a mixture of $^{87}$Rb and K
alkali metals. The cell is heated to about 200$^{\circ }$C by AC currents at
170 kHz in a twisted pair wire heater. The density of Rb at the operating
temperature is measured to be $5.5\times 10^{14}$ cm$^{-3}$, while the
density of K is about $2\times 10^{12}$ cm$^{-3}$.  K atoms are optically
pumped by 650 mW of circularly-polarized light from an amplified distributed Bragg reflector
diode laser at 770 nm. Rb atoms are polarized to 40\% by Rb-K spin exchange
collisions, while $^{21}$Ne atoms are polarized to 15-17\% by spin-exchange
with Rb. The comagnetometer signal is measured by monitoring optical
rotation of a 10 mW linearly polarized probe beam  that is generated by a distributed feedback diode laser near
the 795 nm Rb D1 transition.  The comagnetometer cell is placed inside
magnetic shields consisting of 3 layers of $\mu $-metal and an inner ferrite
shield with an overall shielding factor of $10^{8}$.
Coils inside the magnetic shields are used to cancel residual magnetic
fields and create a compensation  field $B_{z}=-8\pi \kappa
_{0}(M_{Rb}+M_{Ne})/3$, where $M_{Rb}$ and $M_{Ne}$ are the magnetizations of
electron and nuclear spins, and $\kappa _{0}=34\pm 3$ \ is the contact spin-exchange
 enhancement factor for Rb-$^{21}$Ne \cite{Ghosh,Schaefer}. At
this compensation field, the comagnetometer signal is insensitive to
slowly-changing  magnetic fields in all three directions while retaining
sensitivity to anomalous interactions that couple, for example, only to nuclear or to
electron spins. Changes in the sensitivity of the comagnetometer are periodically
monitored  by application of an oscillating $B_{x}$
field \cite{Brown}.   The optical setup is contained in a bell jar evacuated
to 2 Torr to eliminate noise from air currents. The apparatus and all
electronics are mounted on a rotation platform and can be fully rotated around the
vertical axis in several seconds.  To compensate for a shift in the sensitive direction of the co-magnetometer due to an AC Stark shift caused by the 770 nm pump laser at the 780 nm Rb D2 line, we rotate the apparatus by $13^{\circ}$ from the nominal $\hat{y}$ direction toward $\hat{x}$.

A representative sample of the comagnetometer signal during normal data
collection is shown in Fig. 2. The orientation of the platform is reversed by
180$^{\circ }$ every 22 sec. We usually collect data in the North-South (NS)
orientations of the sensitive axis, which gives a maximum
gyroscopic signal due to the  Earth's rotation and in the East-West (EW) orientations,
which gives nominally zero gyroscopic signal. After each mechanical rotation
a background measurement of optical rotation is performed with $B_{z}
$ field detuned by 0.33 $\mu$T to suppress the spin response. Data collection is periodically paused
(300 to 400 sec in Fig. 2) to adjust the compensation value of $B_{z}$ field
and measure the sensitivity of the magnetometer. Fig. 3 shows an example of long-term
measurements of the amplitudes of NS and EW modulations after subtraction of
optical rotation background. We make a 10\% correction to the calibration of the comagnetometer based on the
size of the gyroscopic signal due to Earth's rotation $B_{\rm eff}=\Omega _{\oplus
} I \sin \chi  /\mu _{Ne}=2.63$ pT in the NS data. Slow
drifts in the NS and EW signals are caused by changes in various
experimental parameters. We perform a correlation analysis between the
comagnetometer signal and several measured temperatures, laser beam position
monitors, tilts,  $^{21}$Ne polarization, and other parameters. We find a
small but finite correlation of the signal with oven temperature, room
temperature, and $^{21}$Ne polarization. These correlations are removed from
the data while ensuring that diurnal changes in the temperature do not
accidentally cancel a real sidereal signal in the data.

\begin{figure}[tbp]
\centering
\includegraphics[width=8.5cm]{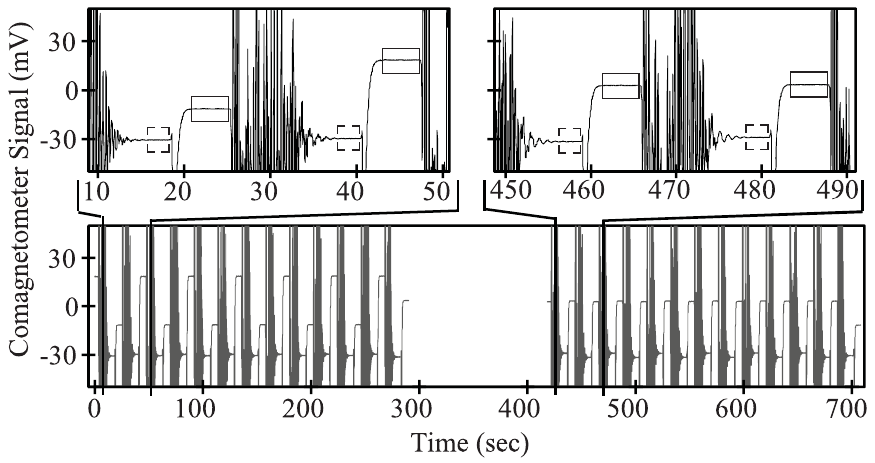}
\caption{Magnetometer signal as a function of time. During the first 300 seconds the sensitive axis is alternated between North and South directions. After the transient from rotation decays, the background
optical rotation is measured (dashed box). Then $B_z$ field is set to the compensation point and actual signal is measured (solid box).
Rotations are paused after 13 reversals for calibration. In the following sequence the sensitive axis is alternated between East and West. }
\end{figure}

\begin{figure}[tbp]
\centering
\includegraphics[width=8.5cm]{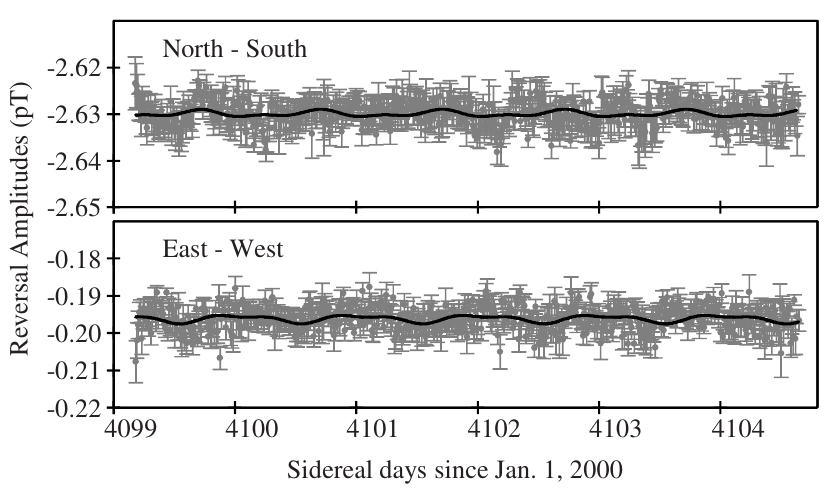}
\caption{Long-term measurements of the N-S and E-W modulation amplitudes with a fit including first and second harmonics of the sidereal frequency. }
\end{figure}

We interpret these measurements within a sub-set of the SME, adding  $c_{\mu \nu }$
coefficients to the standard relativistic Lagrangian of a fermion,
\begin{equation}
\mathcal{L}=\frac{1}{2}i\overline{\psi }(\gamma _{\nu }+c_{\mu \nu }\gamma
^{\mu }) \overleftrightarrow \partial^{\nu }\psi -\overline{\psi }m\psi.
\end{equation}
In the non-relativistic limit, this leads to an anisotropic energy shift for
a particle with spatial components of momentum $p_{j}$, which can be written
as a product of spherical tensor operators of rank 2, $\mathcal{C}_{n}^{2}$
and $\mathcal{P}_{n}^{2}$, formed from Cartesian components of $c_{jk}$ and $%
p_{j}p_{k}$ respectively,
\begin{equation}
H=-c_{jk}p_{j}p_{k}/m=-(-1)^{n}\mathcal{C}_{n}^{2}\mathcal{P}_{-n}^{2}/m.
\end{equation}%
Using the Wigner-Eckart theorem, we evaluate the matrix elements of $\mathcal{P}%
_{n}^{2}\,$\ in terms of $\mathcal{I}_{n}^{2}$, the rank-2 spherical tensor
formed from components of nuclear spin $I,$
\begin{equation}
\left\langle I,m\left\vert \mathcal{P}_{n}^{2}\right\vert I,m^{\prime
}\right\rangle =\frac{\left\langle I,m\left\vert \mathcal{I}_{n}^{2}\right\vert
I,m^{\prime }\right\rangle \left\langle I,I\left\vert \mathcal{P}%
_{0}^{2}\right\vert I,I\right\rangle} {\left\langle I,I\left\vert \mathcal{I}%
_{0}^{2}\right\vert I,I\right\rangle}.
\end{equation}

In the coordinate system defined in Fig.~1, the comagnetometer is sensitive
to first order to the energy of the $^{21}$Ne nuclear magnetic moment $%
\mu _{Ne}$ interacting with the magnetic field in the $\hat{y}$ direction
\cite{Brown},%
\begin{equation}
H_{B}=-\mu _{Ne}I_{y}B_{y}/\left\vert I\right\vert.
\end{equation}
Among tensor nuclear spin operators, the comagnetometer is sensitive to
first order to the operator%
\begin{equation}
H_{Q}=-Q(I_{y}I_{z}+I_{z}I_{y})/2=-iQ(\mathcal{I}_{1}^{2}+\mathcal{I}%
_{-1}^{2})/2,
\end{equation}%
since $I_{z}$ has a finite expectation value due to longitudinal nuclear
polarization of $^{21}$Ne. This gives sensitivity to the $i(\mathcal{C}%
_{1}^{2}+$ $\mathcal{C}_{-1}^{2})$ combination of the spherical tensor
components of the $c_{j k }$ coefficients in the frame of the
experiment. Transforming into the geocentric equatorial
coordinate system, we obtain  expressions for the NS and EW
signals as a function of time,
\begin{eqnarray}
S_{NS} &=&-c_{Y}^{\prime }\cos 2\chi \cos \Omega _{\oplus }t-c_{X}^{\prime
}\cos 2\chi \sin \Omega _{\oplus }t-\nonumber \\
&&(c_{Z}^{\prime }\sin 2\chi \cos 2\Omega
_{\oplus }t+c_{-}^{\prime }\sin 2\chi \cos 2\Omega _{\oplus }t)/2 \nonumber \\
S_{EW} &=&c_{X}^{\prime }\cos \chi \cos \Omega _{\oplus }t-c_{Y}^{\prime
}\cos \chi \sin \Omega _{\oplus }t+\nonumber \\
&&c_{Z}^{\prime }\sin \chi \cos 2\Omega
_{\oplus }t-c_{-}^{\prime }\sin \chi \sin 2\Omega _{\oplus }t,\label{fit}
\end{eqnarray}%
where  the $c$ parameters are defined in Table I  \cite{Kosttable}, $\chi
=49.6^{\circ }$ in Princeton, and $t$ is the local sidereal time. Primes on $%
c$ coefficients indicate that they are measured experimentally in
magnetic field units. The signal contains both first and second harmonics of the Earth rotation rate $\Omega _{\oplus }$.
Each component of the $c$ tensor can be determined independently from NS and EW signals,
although sensitivity to the first harmonic in the NS data is suppressed because $\cos 2\chi=-0.16$.
Vector Lorentz-violating interactions, such as the $b_{\mu }$ coefficient, can
contribute to the first harmonic signal, however within the Schmidt nuclear
model for $^{21}$Ne with a valence neutron such an interaction has
already been excluded with sufficient precision by previous experiments
with $^{3}$He \cite{Brown} and electrons \cite{Heckel}. The results of the daily fits for the $c$
coefficients are shown in Fig. 4 for our data spanning a period of 3 months
and the final results are summarized in Table I. The systematic uncertainty
is determined by the scatter of several analysis procedures, including fits
of data without removal of long-term correlations and varying data cuts.

\begin{figure}[tbp]
\centering
\includegraphics[width=8.5cm]{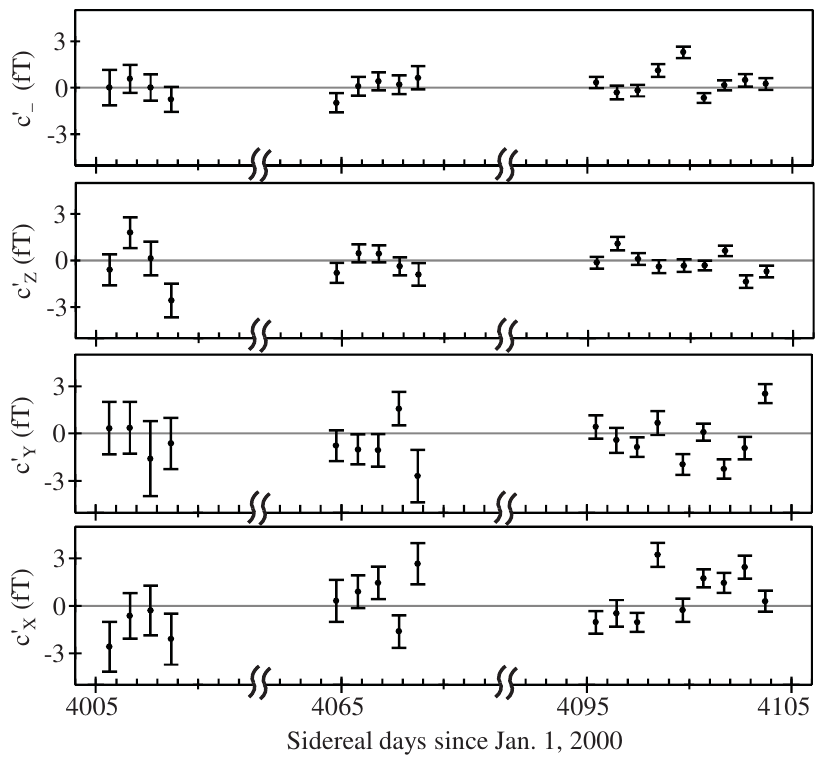}
\caption{Daily fits for the $c$ coefficients defined in Eq.~(\ref{fit}). The final statistical uncertainty is increased
to account for reduced $\chi^2=2$ to $4$. }
\end{figure}

\begin{table*}[t]
\begin{ruledtabular}
\begin{tabular}{lcccc}
& North-South (fT) & East-West (fT) & Combined (fT) & Scaled   \\ \hline
$c_{X}=c^n_{YZ}+c^n_{ZY}$ & $-0.9\pm 1.8\pm 2.3$ & $0.75\pm 0.39\pm 0.50$ & $%
0.67\pm 0.62$ & $(4.8\pm 4.4) \times 10^{-29}$ \\
$c_{Y}=c^n_{XZ}+c^n_{ZX}$ & $0.7\pm 1.3\pm 2.5$ & $-0.42\pm 0.36\pm 0.33$ & $%
-0.39\pm 0.48$ & $-(2.8\pm 3.4) \times 10^{-29}$ \\
$c_{Z}=c^n_{XY}+c^n_{YX}$ & $0.01\pm 0.34\pm 0.27$ & $-0.23\pm 0.19\pm 0.13$ & $%
-0.17\pm 0.20$ & $-(1.2\pm 1.4) \times 10^{-29}$\\
$c_{-}=c^n_{XX}-c^n_{YY}$  & $0.65\pm 0.34\pm 0.32$ & $0.04\pm 0.19\pm 0.20$ & $%
0.20\pm 0.24$ & $(1.4\pm 1.7) \times 10^{-29}$%
\end{tabular}
\end{ruledtabular}
\caption{Results for the neutron $c^n_{jk}$ coefficients measured in $^{21}$Ne.}
\end{table*}

To convert our measurements from magnetic field units, we need an estimate of
the nuclear operator $\left\langle I,I\left\vert \mathcal{P}%
_{0}^{2}\right\vert I,I\right\rangle =\left\langle I,I\left\vert
2p_{z}^{2}-p_{x}^{2}-p_{y}^{2}\right\vert I,I\right\rangle /\sqrt{6}$.
Within the Schmidt model, $^{21}$Ne has a valence neutron in the $%
d_{3/2}$ state. This gives $\left\langle I,I\left\vert \mathcal{P}%
_{0}^{2}\right\vert I,I\right\rangle =-\sqrt{2/3}\left\langle
p^{2}\right\rangle /5=-\sqrt{8/3}mE_{k}/5$ and we take the kinetic energy of
the valence nucleon $E_{k}\sim 5$ MeV \cite{clockcom}. $^{21}$Ne is better described by a
collective wavefunction within the $sd$ shell model, and it should be possible to calculate the nuclear operator
more precisely \cite{Wilbert}.

To evaluate the relative sensitivity of the comagnetometer to $H_{B}$ and $%
H_{Q}$, we model the equilibrium spin density matrix of $^{21}$Ne. Two
processes dominate the spin evolution of $^{21}$Ne, binary spin-exchange
collisions with Rb \cite{Happer84} and spin-relaxation due to the interaction of
its nuclear electric quadrupole moment with electric field gradients during
binary atomic collisions \cite{CohTan}. At low Rb density, we measured $^{21}$%
Ne spin relaxation time constant $T_{Q}=$ $87\pm 4~$min, close to the
relaxation rate of $103\pm 6$ min that is predicted for our $^{21}$Ne \
density from previous measurements of the $^{21}$Ne nuclear quadrupole
relaxation rate \cite{Ghosh}. At normal operating temperature,  the $^{21}
$Ne spin time constant was equal to $63\pm 2~$min, implying an additional Rb-%
$^{21}$Ne spin-exchange time constant $T_{ex}=230\pm 40$ min. This is
reasonably consistent with Rb-$^{21}$Ne spin-exchange time constant of $%
370\pm 70$ min calculated from a previously measured Rb-$^{21}$Ne spin-exchange
rate constant \cite{Ghosh}. The equilibrium $^{21}$Ne polarization of 17\%
is also consistent with measured Rb polarization and spin-exchange and
relaxation rates. In the density matrix model for the comagnetometer, we
find the ratio of the signals produced by $H_{B}$ and $H_{Q}$,%
\begin{equation}
\frac{S(Q)}{S(B_{y})}=\frac{Q\left\langle I_{z}\right\rangle }{2\mu
_{Ne}B_{y}/I }f(T_{ex}/T_{Q})
\end{equation}%
where the function $f(T_{ex}/T_{Q})$ ranges from $f(0)=1$ to $f(\infty )=3/2$
as the population distribution \ among the four spin states of $^{21}$Ne
changes depending on the nature of the dominant spin relaxation mechanism.
For our operating parameters $f(2.6\pm 0.4)=1.27\pm 0.03$. The conversion
factor to dimensionless $c$ coefficients parameterizing variations in the neutron MAV is %
\begin{equation}
c=-c^{\prime }\frac{10\mu _{n}}{3P_{Ne}fE_{k}}=7.1\times 10^{-29}
\end{equation}%
and our results for the neutron $c^n_{jk}$ are given in the last column of Table I.

Our limits are 2 to 4 orders of magnitude more stringent than
previous limits set with similar experiments using $^{7}$Be \cite{Prestage}, $^{201}$Hg \cite{Lamoreaux}, and $^{21}$Ne \cite{Chupp} spins. They are also 4 orders of magnitude
more stringent than limits on similar proton $c$ coefficients set using $%
^{133}$Cs atomic fountain clock \cite{Wolf} and many orders of magnitude more
sensitive than limits for electron $c$ coefficients \cite{Kosttable}.  Among
astrophysical tests, the most stringent limits have been set on $c_{0j}$ for
neutrinos by Icecube at $10^{-27}$ \cite{Icecube} and on $c_{00}$ for protons at $10^{-23}$ from
analysis of the spectrum of ultra-high energy cosmic rays \cite{Scully,Bi}. Our
measurements of the spatial components $c_{ij}$  are comparable to these results for $c_{00}$ and $c_{0j}$ if we assume that Lorentz violation is
manifested by a difference between space and time in a preferred frame of the Cosmic Microwave Background, which moves relative to the Earth with a
velocity of $10^{-3}c$ in an approximately equatorial direction.

This measurement represents the first experimental result using the $^{21}$Ne-Rb-K
comagnetometer. Our frequency resolution is similar to the previous K-$^{3}$%
He Lorentz-violation experiment \cite{Brown} with a factor of 8 shorter integration time
and without full optimization of the apparatus. The shot-noise sensitivity
of the comagnetometer is already more than an order of magnitude better
than the best results obtained with K-$^{3}$He \cite{Vasilakis}, allowing
for further improvements in both vector and tensor Lorentz invariance tests as
well as other precision measurements by several orders of magnitude. This work was supported by NSF Grant No. PHY-0969862.

\end{document}